# Tunable Thermal Energy Transport across Diamond Membranes and Diamond-Si Interfaces by Nanoscale Graphoepitaxy

Zhe Cheng[1], Tingyu Bai[2], Jingjing Shi[1], Tianli Feng[3, 4], Yekan Wang[2], Matthew Mecklenburg[2], Chao Li[2], Karl D. Hobart[5, *], Tatyana I. Feygelson[5], Marko J. Tadjer[5], Bradford B. Pate[5], Brian M. Foley[1], Luke Yates[1], Sokrates T. Pantelides[3, 4], Baratunde A. Cola[1, 6], Mark Goorsky[2], Samuel Graham[1, 6, *]

[1] George W. Woodruff School of Mechanical Engineering, Georgia Institute of Technology, Atlanta, Georgia 30332, USA

[2] Materials Science and Engineering, University of California, Los Angeles, Los Angeles, CA, 91355, USA

[3] Department of Physics and Astronomy and Department of Electrical Engineering and Computer Science, Vanderbilt University, Nashville, Tennessee 37235, USA

[4] Materials Science and Technology Division, Oak Ridge National Laboratory, Oak Ridge, Tennessee 37831, USA

[5] U.S. Naval Research Laboratory, 4555 Overlook Ave SW, Washington, DC 20375, USA

[6] School of Materials Science and Engineering, Georgia Institute of Technology, Atlanta, Georgia 30332, USA

[*] Corresponding authors: sgraham@gatech.edu; karl.hobart@nrl.navy.mil

**ABSTRACT**

The development of electronic devices, especially those that involve heterogeneous integration of materials, has led to increased challenges in addressing their thermal operational-temperature demands. The heat flow in these systems is significantly influenced or even dominated by thermal boundary resistance at interface between dissimilar materials. However, controlling and tuning heat transport across an interface and in the adjacent materials has so far drawn limited attention. In this work, we grow chemical-vapor-deposited (CVD) diamond on silicon substrates by graphoepitaxy and experimentally demonstrate tunable thermal transport across diamond membranes and diamond-silicon interfaces. We observed the highest diamond-silicon thermal boundary conductance (TBC) measured to date and increased diamond thermal conductivity due to strong grain texturing in the diamond near the interface. Additionally, non-equilibrium molecular-dynamics (NEMD) simulations and a Landauer approach are used to understand the diamond-silicon TBC. These findings pave the way for tuning or increasing thermal conductance in heterogeneously integrated electronics that involve polycrystalline materials and will impact applications including electronics thermal management and diamond growth.

## 1. INTRODUCTION

The ongoing miniaturization of microelectronic devices, as well as their heterogeneous integration to create advanced functionalities, have led to high local power densities and circumstances where thermal effects limit the overall device performance.[1-3] Keeping these devices cool has become a design challenge aiming to avoid the degradation of device performance and reliability.[2-3] Due to the architecture of these electronic systems, heat dissipation can be significantly influenced or even dominated by the thermal boundary resistance found at heterointerfaces.[4-5] Previous efforts to reduce thermal boundary resistance between solids include bridging phonon spectra mismatch and enhancing interfacial bonding.[6-12] In addition, several theoretical studies show that incorporating nanostructures at the interface enlarges the interface contact area and increases TBC, but experimental results are inconsistent.[13-17] Tuning thermal transport across interfaces or even in the adjacent materials remains largely an open issue.

Graphoepitaxy is a technique that uses artificial surface relief structures to induce crystallographic orientation in thin films grown on a surface.[18-21] This technique was invented to grow Si, Ge, and KCl on amorphous $SiO_2$ substrates about four decades ago.[18-21] After that, it was extensively used to grow block copolymers and carbon nanotubes to control orientation or alignment.[22-24] By introducing nanoscale graphoepitaxy into thermal transport across interfaces, the solid-solid interface contact area increases due to the artificial surface structures, which may contribute to increasing TBC. The crystallographic orientation of grains in the adjacent membranes may affect their thermal conductivity as well. These two synergistic effects provide a possible solution to tune thermal transport across interfaces and in the adjacent membranes.

In this work, we successfully grow diamond membranes on silicon substrates by nanoscale graphoepitaxy. Time-domain thermoreflectance (TDTR) is used to measure the thermal conductivity of the diamond layer and the diamond-silicon TBC. The diamond thermal conductivity and diamond-silicon TBC are tuned with different surface pattern sizes. Scanning transmission electron microscopy (STEM) and X-ray diffraction (XRD) are used to study the grain size distribution and orientation. NEMD simulation and a Landauer approach are used to understand the diamond-silicon TBC. Our work is notably the first effort to tune diamond growth on silicon substrates, and subsequently thermal transport across diamond-silicon interfaces and diamond membranes by graphoepitaxy. We expect that graphoepitaxy can be applied to polycrystalline diamond grown on other substrates as well.

## 2. SAMPLES AND METHODOLOGIES

**2.1. Samples.** In this work, six silicon wafers are prepared (Samples A1, B1, ref1, and A2, B2, ref2). Samples A1, A2, B1, and B2 are patterned silicon wafers with nanoscale trenches while Samples ref1 and ref2 are flat silicon wafers without nanoscale trenches. The dimensions of the interface patterns are summarized in Table 1. For example, scanning electron microscopy (SEM) images of the patterned Si trenches of Sample A2 are shown in Figure S1. Nanocrystalline diamond (NCD) films were fabricated with the same growth conditions on both nanopatterned and flat silicon substrates acquired from LightSmyth Technologies. NCD was grown on a flat (100) oriented polished silicon substrate by a microwave plasma-assisted chemical vapor deposition (MPCVD) method in IPLAS 5.0 KW CVD reactor with hydrogen and methane as reactant gases. The growth conditions were consistent throughout the entire deposition process as follows: 750 °C substrate temperature, 7.0 Torr chamber pressure, 1400 W microwave power, and 0.5% methane to hydrogen ratio. The flat Si substrate enables in-situ NCD film thickness measurement using laser reflectometry, and also serve as a reference for the future comparison with the patterned silicon. Prior to diamond growth, all the silicon substrates were seeded by ultrasonic treatment in ethanol-based nanodiamond suspension prepared from detonation nanodiamond powder (International Technology Center, North Carolina, USA (ITC)). According to the manufacturer specifications the material grade used here has a high degree of grain size homogeneity with an average particle size of 4 nm, and a chemical purity in excess of 98%. The SEM analysis of the backside of a typical NCD film deposited with implementation of the abovementioned seeding method shows a uniform seed density greater than $10^{12}$ nuclei/cm$^2$. With this type of diamond nucleation, the NCD films were formed through grain coalescence and subsequent growth competition of initially random-oriented nanodiamond seeds. Only the crystals with the fastest growth speed along the thickness direction extend to the surface. This process ultimately leads to a formation of a well-pronounced columnar grain structure in the film as well as an increase in lateral grain size with film thickness. The use of carbon-lean growth conditions as above is intended to suppress secondary renucleation and increases film quality by reducing grain boundaries.

Table 1. Dimensions of Si patterns for Samples A, B, and ref.

| Sample | Height | Top width | Bottom width |
|--------|--------|-----------|--------------|
|        | nm     | nm        | nm           |
| A      | 47     | 60        | 77           |
| B      | 105    | 205       | 215          |
| ref    | 0      | 0         | 0            |

Samples A1, B1 and ref1 were used for material characterization with 1-um-thick diamond films. Samples A2, B2 and ref2 were used for TDTR measurements with 2-um-thick diamond films to improve TDTR sensitivity. Thermal properties of the diamond films and diamond-silicon interfaces were obtained from the TDTR measurements. All the diamond layers were grown under the same conditions.

**2.2. Thermal Characterization.** The thermal properties in this work are measured by multi-frequency TDTR.[25-29] TDTR is a well-established noncontact optical pump-and-probe thermal characterization tool used to measure thermal properties of both bulk and nanostructured materials.[28, 30-31] As shown in Figure 1, a pump beam which is chopped by a modulator heats a sample periodically and a delayed probe beam measures the temperature decay of the sample surface through a change in thermoreflectance. The probe beam delay time is controlled by a mechanical stage, which is used to create a temperature decay curve from 0.1 to 5 ns. By fitting the experimental signal picked up by a lock-in amplifier with an analytical solution of heat flow in the layered structure, one or more thermal properties of the sample can be extracted.[25-29, 31] In TDTR measurements, the distance that the heat penetrates into the surface depends on the modulation frequency and the thermal diffusivity of the sample. By tuning the modulation frequency, we infer the thermal properties of the sample with different penetration depths, leading to different sensitivity to different unknown parameters. If we measure one spot on a sample with different modulation frequencies, we obtain TDTR data that are sensitive to more than one unknown parameters. By fitting these TDTR curves simultaneously, we obtain the values of these unknown parameters. The definition of TDTR sensitivity is shown in Equation S1 and the sensitivity analysis of the multi-frequency TDTR measurements can be found in Figure S2.

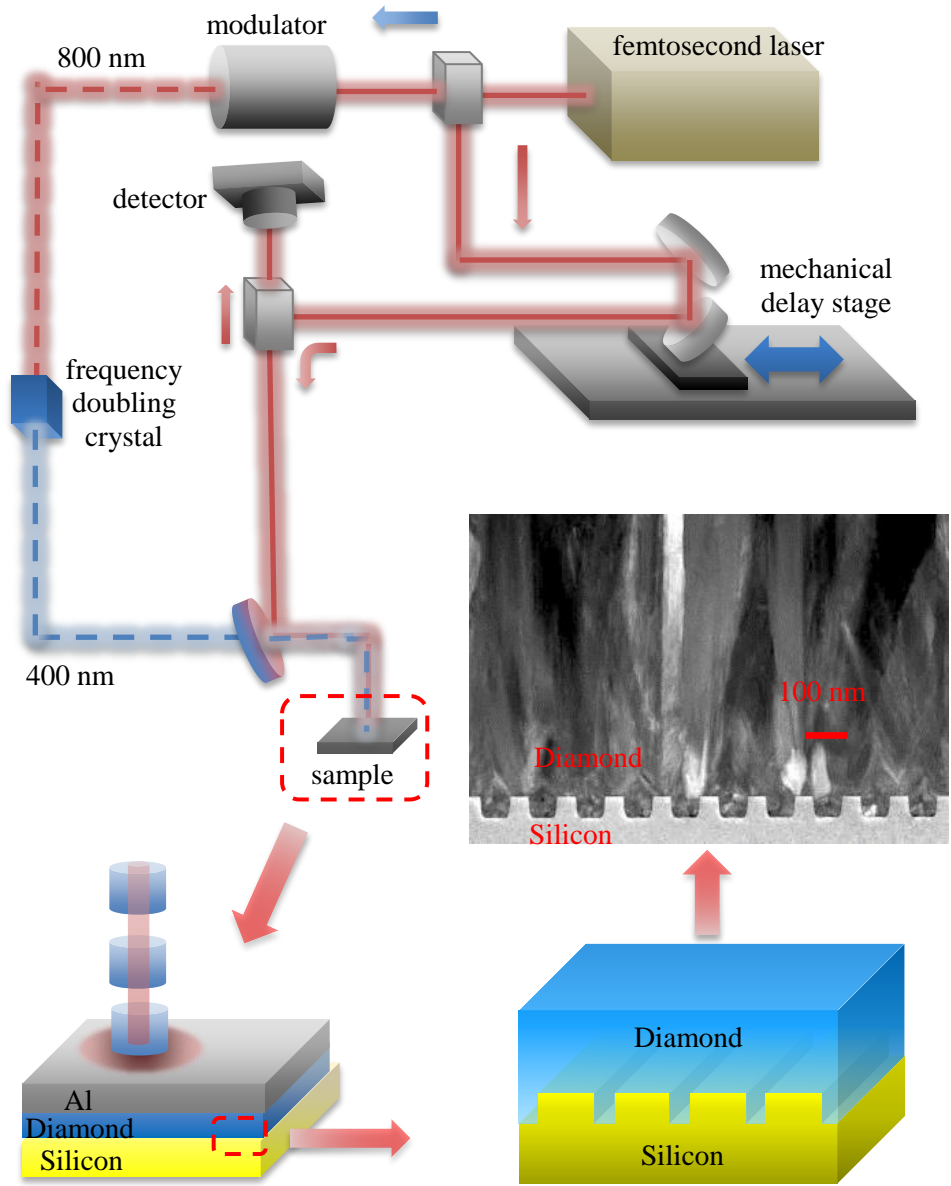

Figure 1. Schematic diagram of TDTR and sample structure grown by graphoepitaxy with nanoscale patterns. The TEM image shows the patterned diamond-silicon interface (CVD diamond grown on patterned silicon substrates by graphoepitaxy).

To perform TDTR measurements, a layer of aluminum (Al) is deposited on the sample surface as a transducer. The Al thicknesses are determined by the picosecond acoustic method[32-33] (those of

Samples A2, B2, and ref2 are 103 nm, 80 nm, and 74 nm, respectively). The thermal conductivity of the Al layer is determined by measuring its electrical conductivity and applying the Wiedemann-Franz law. The thermal conductivity of the silicon substrate is taken from the literature (142 W/m-K).[34] The thickness of the diamond layers in Samples A2, B2, and ref2 are measured to be 2.3 µm by a SEM. The density and specific heat of CVD diamond and Al used for the analysis of the data are from the literature.[28] The pump and probe beam size (radius) are 8.1 µm and 6.4 µm for Samples A2, B2. Those of Sample ref2 are 7.7 µm and 7.5 µm, measured with a DataRay scanning slit beam profiler. A standard silicon calibration sample is checked every time before measuring the diamond samples. Three-frequency TDTR measurements are used to measure the Al-diamond TBC, diamond cross-plane thermal conductivity, and diamond-silicon TBC. As shown in Figure S2, the TDTR signal is more sensitive to diamond cross-plane thermal conductivity at high modulation frequency because heat does not penetrate through the diamond-silicon interface. The TDTR signal is more sensitive to diamond-silicon TBC at low modulation frequency because heat penetrates deep into the silicon substrate. Therefore, 1.2-3.6 MHz or 2.2-6.3 MHz can be chosen to perform the three-frequency TDTR measurements. An example of good agreement of multi-frequency TDTR data fitting of theoretical curves and experimental data is shown in Figure S3.

**2.3. Material Characterization.** Plan-view and cross-section TEM samples were prepared using Focused Ion Beam (Nova 600 SEM/FIB). The schematic diagram of TEM sample preparation with FIB can be found in Figure S4. The near-interface plan-view samples were made at the Si patterned region so that both silicon and diamond can be seen. STEM images were then generated using a Titan S/TEM (FEI) system under 200 kV. The STEM mode with a high-angle annular dark-field (HAADF) detector provides images with contrast due to differences in the adjacent grain orientation. The cross-section STEM images were used to study the grain growth near the nucleation region. The plan-view images were used to measure average grain size and its distribution within an area. Dark-field images were also taken to show grains with either (111) or (110) plane parallel to the sample surface. These images were used to calculate the grain growth ratio for (111) and (110) oriented grains (more details can be found later). X-ray diffraction (XRD) was used to analyze cross-plane preferred grain orientation. The XRD 2θ:ω scan was performed on a Jordan Valley D1 diffractometer with Cu K$\alpha_1$ radiation and a parallel beam source. In these measurements, ω was offset by 5-10° from the surface orientation of Si substrate to avoid the strong (004) Si reflection. This offset won't influence the measurement of the preferred orientation.

**2.4. NEMD Simulations.** The MD simulations were performed using the LAMMPS[35] code and Tersoff potentials.[36] The simulation domain contains a 230-Å-long diamond (28,080 atoms) and a 330-Å-long Si (11,712 atoms) with the same cross-section area of 32.8*21.6 Å$^2$. The temperature difference is applied along the x direction, and periodic boundary conditions are applied along y and z directions. In the NEMD simulations, the domains were first stabilized at 300 K by NPT simulations (constant pressure and temperature) with 2,000,000 steps and then converted to NVE (constant volume and energy) ensemble, with the temperatures of 350 K and 250 K applied at the ends of diamond and Si, respectively. 3,000,000 steps of NVE simulations were used to stabilize the temperature gradient and heat current through the whole system. The time for each step is 0.5 fs. After that, another 2,000,000 NVE steps simulations were performed to extract the stabilized temperature gradient and heat flux. The amorphous layer was constructed before the NEMD simulations by melting the 20 Å-long region of diamond at the interface at 3000 K with fixed volume (20% larger than crystalline diamond to allow atoms to move), followed by an annealing process to 300 K at a rate of 0.54 K/ps (10,000,000 steps) as well as a NPT relaxation.

**2.5. Landauer Approach.** The Landauer approach is a widely used method to predict TBC ($G$)[4, 37-41] and it has been applied here to calculate the TBC at the diamond-silicon interface. The general form of the Landauer formula calculating $G$ at a 3D/3D interface is:

$$G = \frac{q}{A\Delta T} = \frac{1}{A\Delta T} \left( \sum_p \frac{A}{2} \iint D_1(\omega) f_{BE}(T_1) \hbar \omega v_1(\omega) \tau_{12}(\theta, \omega) \cos\theta \sin\theta \, d\theta d\omega - \sum_p \frac{A}{2} \iint D_2(\omega) f_{BE}(T_2) \hbar \omega v_2(\omega) \tau_{21}(\theta, \omega) \cos\theta \sin\theta \, d\theta d\omega \right), \quad (1)$$

where $q$ is the net heat flow rate, $A$ is the cross-sectional area of the interface, $D$ is the phonon density of states (DOS), $f_{BE}$ is the Bose-Einstein distribution function, $\hbar$ is the reduced Planck constant, $\omega$ is the phonon angular frequency, $v$ is the phonon group velocity, $\tau_{12}$ is the transmission coefficient from material 1 to 2 (here it is from silicon to diamond), $\theta$ is the angle of incidence, and the sum is over all phonon modes. With the restriction of detailed balance, the formula can be simplified as:

$$G = \frac{q}{A\Delta T} = \frac{\sum_p \frac{1}{2} \iint D_1(\omega) \left( f_{BE}(T_1) - f_{BE}(T_2) \right) \hbar \omega v_1(\omega) \tau_{12}(\theta, \omega) \cos\theta \sin\theta \, d\theta d\omega}{\Delta T}. \quad (2)$$

Without considering the local non-equilibrium near the interface, the formula can be further simplified as:

$$G = \sum_p \frac{1}{2} \iint D_1(\omega) \frac{df_{BE}}{dT} \hbar \omega v_1(\omega) \tau_{12}(\theta, \omega) \cos\theta \sin\theta \, d\theta d\omega. \quad (3)$$

Here, we use the diffuse mismatch model (DMM) to calculate the transmission coefficient.[38-39, 41]

$$\tau_{12}(\omega) = \frac{\sum_p M_2(\omega)}{\sum_p M_1(\omega) + \sum_p M_2(\omega)}, \tag{4}$$

where $M$ is the number of modes, which is proportional to the square of the wave vector for a 3D isotropic material. The DMM assumes all the incident phonons are diffusely scattered at the interface and lose their memory.

## 3. RESULTS AND DISCUSSION

### 3.1. Enhanced Thermal Transport across Interfaces

To dissipate the localized Joule heating in power electronics, CVD diamond is an excellent candidate for thermal management because of its high thermal conductivity.[42-46] However, when integrating diamond with other materials, the TBC is very small due to the large mismatch in phonon DOS between diamond and other materials. Generally speaking, phonons with a certain frequency have a high likelihood to transmit through an interface only when phonons with this frequency exist on the other side of the interface or when specific modes that are local to the interface help the transmission of those phonons.[47-52] Therefore, the degree of DOS overlap between two adjacent materials has a significant effect on the TBC across an interface. Due to the small mass of carbon atoms and strong bonds among these carbon atoms in diamond, diamond has a very high cutoff frequency (the Debye temperature of diamond is 2230 K).[53] When integrating diamond with other materials, poor DOS overlap and a correspondingly small TBC are expected. Figure 2(a) shows a comparison of the DOS of diamond and several typical materials (Pt, MgO, SiC, and Si). The DOS overlaps between diamond and these materials are small, leading to small TBC.[54-55]

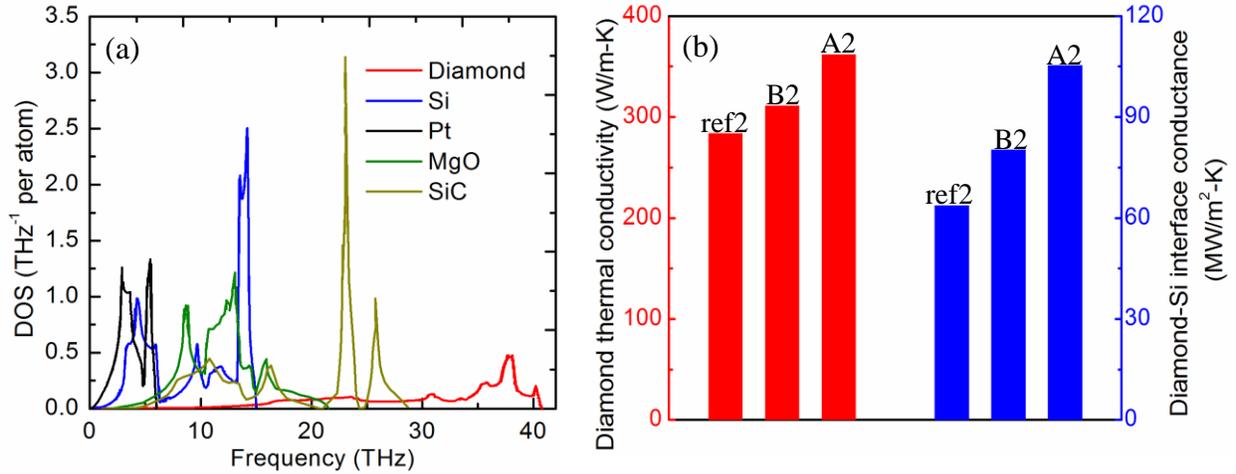

Figure 2. (a) Phonon DOS of diamond and a few other materials, highlighting the sizable differences in the vibrational spectra of different crystalline materials.[54-56] (b) Comparison of the cross-plane thermal conductivity of diamond layers and diamond-silicon TBC for the flat sample (ref2) and the patterned samples (A2 and B2).

By using multi-frequency TDTR measurements,[25-29, 31] we measured the diamond cross-plane thermal conductivity and the diamond-silicon TBC at room temperature and the results are shown in Figure 2(b). Here, we will discuss the TBC of flat diamond-silicon interfaces first. The TBC of the flat interface in this work is measured to be 63.7 MW/m$^2$-K, which is very close to the value measured by Joule-heating method in the work done by Goodson *et al*. ( about 66.7 MW/m$^2$-K)[57-58], larger than the value measured by 3-Omega method by Mohr *et al*. (50 MW/m$^2$-K)[59]. These measured TBC of flat diamond-silicon interfaces from literature and this work are close to 60 MW/m$^2$-K and generally agree with each other.

In terms of theoretical calculations and simulations for diamond-silicon TBC, Khosravian *et al*. calculated the diamond-silicon TBC using NEMD. The TBC is determined as 335.6 MW/m$^2$-K, which is 5 times larger than our measured value.[60] We used NEMD to calculate the diamond-silicon TBC as well. The TBC is found to be 381 MW/m$^2$-K. The difference between these NEMD results derives from the difference of the used atomic potentials and size effects of finite simulation domains. The calculated TBC from the Landauer formula with transmission from DMM is 316.9 MW/m$^2$K. These theoretical values calculated by NEMD and the Landauer approach with DMM are close to 350 MW/m$^2$-K and generally agree with each other while they are much larger than the experimental values. We attribute the difference between the experimental results and the theoretical values to interface bonding. The calculated TBC values are based on perfect diamond-silicon interfaces with covalent bonding while the measured interfaces may have weak interface bonding, such as Van der Waals force bonding, for some area of the interface.

Now we turn to our measured TBC of the nanopatterned interfaces. The measured diamond-silicon TBC for the sample grown by graphoepitaxy (sample A2) is 105 MW/m$^2$-K, which is the highest diamond-silicon TBC measured to date. We attribute this high measured TBC to enlarged contact area between diamond and silicon. When comparing with the flat diamond-silicon interface, the diamond-silicon TBC of A2 increases by 65% for the nanopatterned interface. The patterned interface enlarges the diamond-silicon contact area, which behaves like fins in convective heat transfer. Because the fin length is very short, the relation between the ratio of the TBC and the ratio of contact area should be as below:

$$\frac{G_p}{G_{ref}} \approx \frac{S_p}{S_{ref}} \tag{5}$$

Here, $G_p$ and $G_{ref}$, $S_p$ and $S_{ref}$ are the TBC and contact areas of the patterned and reference samples. $S_p = L_t + L_b + 2h$ and $S_{ref} = L_t + L_b$. Here, $L_t, L_b, h$ are the top width, bottom width, and height of the pattern. The contact area of the patterned interface (Samples A1 and A2) increases by 69% ($S_p/S_{ref} - 1$) compared with that of the flat diamond-silicon interface (Samples ref1 and ref2). This consistency between TBC enhancement (65%) and contact area enlargement (69%) confirms that the increased TBC is due to the larger contact area. Here we experimentally confirm the effect of increased contact area on TBC predicted by the theoretical calculations and simulation works in the literature.[13-15] For Sample B2, the TBC is also enhanced by 26%, but it is smaller than the contact area enhancement (50%). This difference may be due to the grain impingement we will discuss later, which facilitates good contact between the diamond and the side walls of the silicon patterns.

To explore the mechanism behind the enhanced thermal conductance across the interface, STEM and XRD are used to characterize the structure of the diamond-silicon interfaces. The STEM images in Figure 3(a-b) were taken using the HAADF detector to show the contrast from different grains. They show that the grains nucleating from the silicon surface tend to impinge upon one another, coalescing together in the area located above the trenches. Figure 3(a) shows a plan-view STEM image that includes the diamond-silicon interface and Figure 3(b) shows a cross-section STEM image of the diamond-silicon interface. We can clearly see the patterned silicon ridges in Figure 3(a). The diamond grains grow on the Si trench and eventually impinge at the middle of the trench region, as indicated by the yellow dashed lines in Figure 3(a-b). This grain impingement affects the preferred crystal orientation and corresponding thermal properties. First, the grain impingement forces the grown diamond to have very good contact with the silicon nanoscale trenches. We do not observe any voids near the interface. This good contact facilitates thermal transport across the interface and enhances the TBC. This may be the reason that the TBC enhancement of Sample A2 matches well with contact area enhancement. Second, the grain impingement induces preferred grain orientation (texturing) in the continually grown diamond layer.

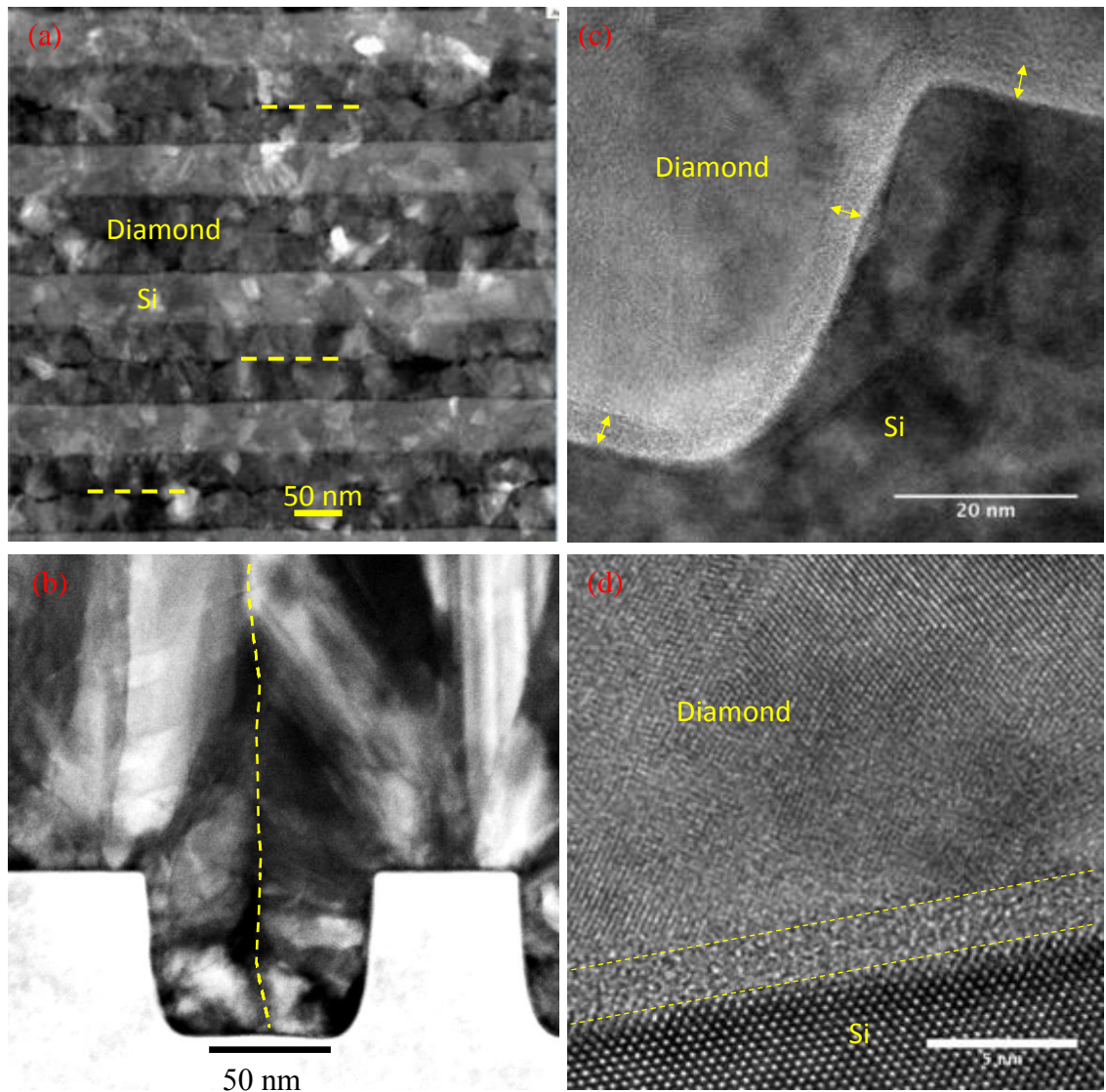

Figure 3. Grains impinge over the patterned trenches (Sample A1) and amorphous layer at the diamond-silicon interface. (a) Plan-view STEM image near the diamond-silicon interface. (b) Cross-section STEM image of diamond-silicon interface. (c-d) Cross-section HRTEM images to show the amorphous carbon region at the diamond-silicon interfaces of the patterned and flat samples.

Figure 3(c-d) includes high resolution transmission electron microscopy (HRTEM) images taken at the diamond-silicon interfaces showing lattice fringes for the silicon substrate and diamond grains. As shown in Figure 3(c-d), no SiC is observed at or near the interfaces for either the patterned or the flat Si-diamond interfaces. However, an amorphous layer is present (about 2 nm thick) for both interfaces. Electron energy loss spectroscopy (EELS) is performed on the flat interface and the results are shown in Figure 4. The measurements were performed in four regions including the pure diamond region, the diamond-silicon interface region and the silicon substrate region. The EELS results show the existence of $\leq 4$ nm (measurement resolution) $sp^2$ C at the interface. The EELS measurement combined with the HRTEM image supports the conclusion that a 2-nm amorphous layer observed in the HRTEM is $sp^2$ C, which is formed during the diamond deposition process for both patterned and flat samples.

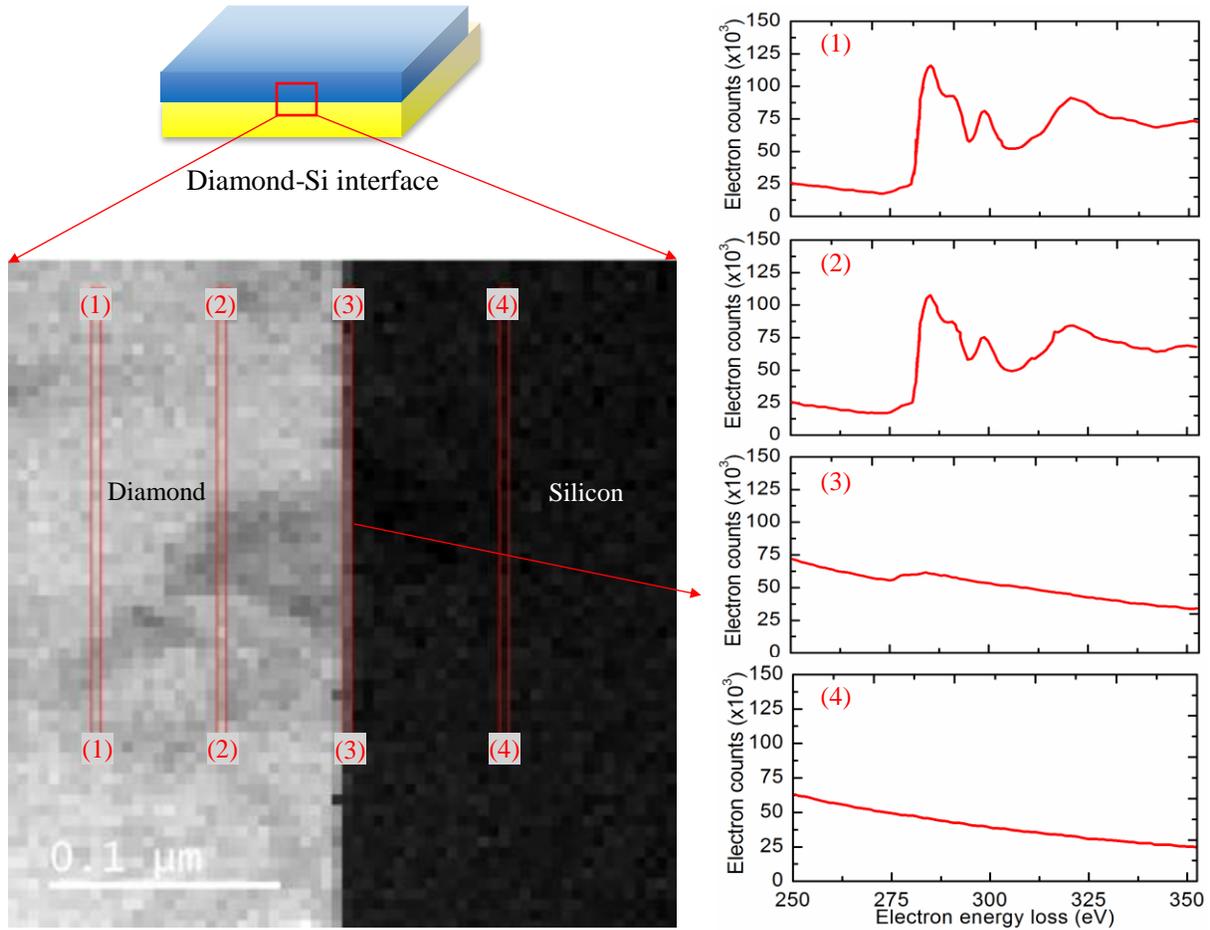

Figure 4. EELS data of diamond-silicon interface. The measurements were performed in four regions including the pure diamond region (1-2), the diamond-Si interface region (3), and the Si substrate region (4). The results show the existence of < 4 nm (length of 1 pixel) $sp^2$ C at the interface.

To study the effect of amorphous carbon at the diamond-silicon interface on thermal transport, we performed NEMD simulations. As shown in Figure 5, a temperature difference is applied across the diamond-silicon interface. We find that the interface between amorphous diamond and silicon presents larger thermal conductance than that between crystalline diamond and silicon, i.e., the temperature jump at the interface (x=22 Å) becomes smaller after the amorphous diamond layer is introduced. This is consistent with Si-Ge interfaces in the literature.[61] The TBC of an interface involving amorphous materials may be slightly larger than that of an interface only involving their crystalline counterparts. For our work, the existence of the amorphous carbon increases the diamond-silicon TBC while the amorphous carbon layer itself has thermal resistance so the overall interfacial thermal conductance does not change much, i.e., the total temperature jump does not change much since the amorphous diamond layer introduces an extra temperature jump as seen from x=20 to 22 Å. This effect arises because the amorphous diamond has a lower thermal conductivity than crystalline diamond. The overall TBC is determined as 381 MW/m$^2$-K without amorphous carbon and 378 MW/m$^2$-K with amorphous carbon for the systems, close to the previous TBC value calculated by NEMD.[60] Overall, the effect of the amorphous layer on the diamond-silicon TBC is negligible (smaller than 1%). The intrinsic diamond-silicon thermal boundary resistance is the dominant thermal resistance at the interfaces. Since this amorphous carbon layer exists for both the flat and patterned samples, the effect should be the same for both interfaces. Therefore, the existence of the amorphous layer does not affect our conclusion that the enhanced thermal transport across the diamond-silicon interface grown by graphoepitaxy is due to the enlarged contact area.

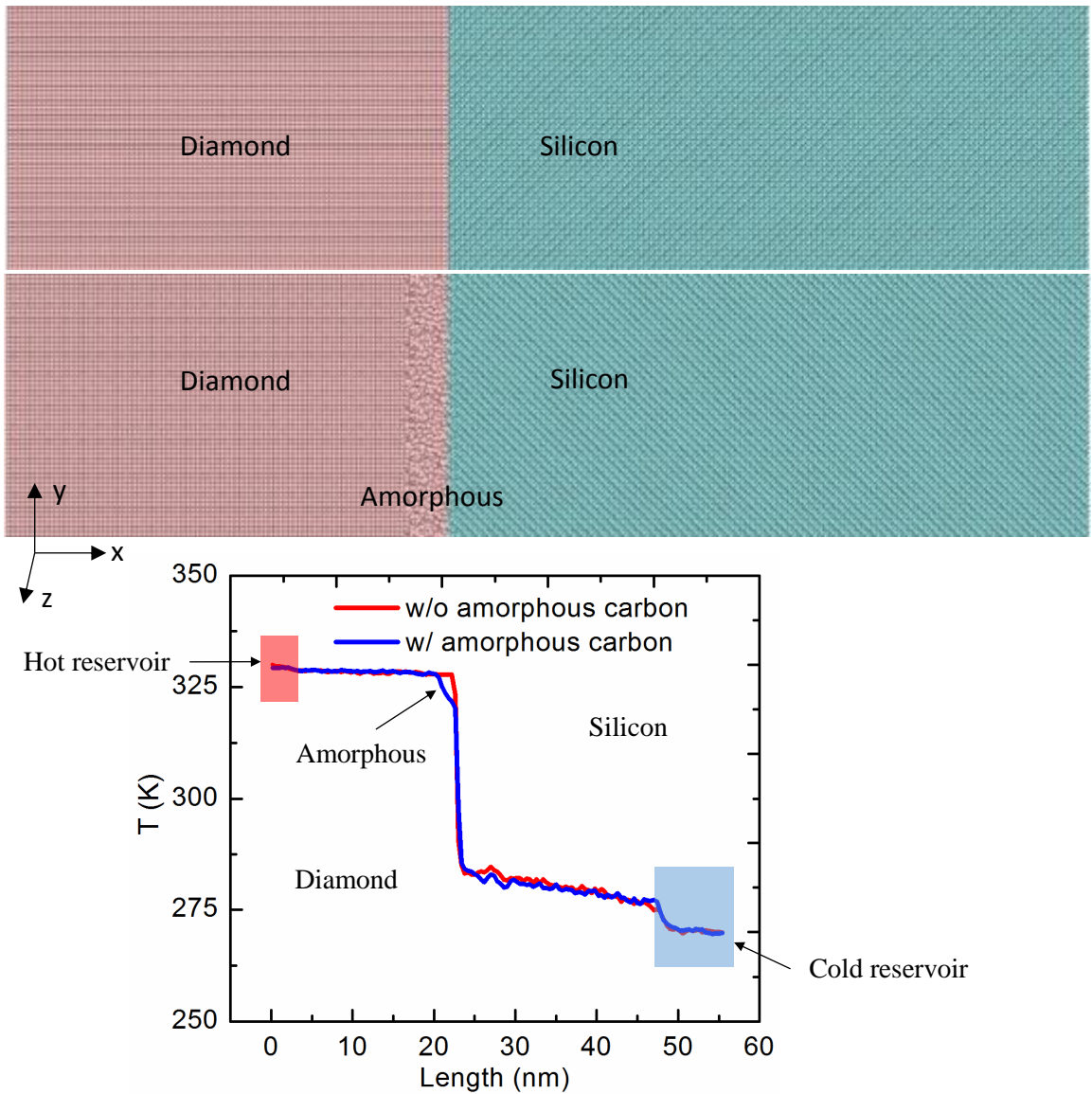

Figure 5. NEMD simulation of thermal transport across the diamond-silicon interfaces with and without amorphous carbon layer. The effect of the amorphous layer on diamond-silicon TBC is negligible (<1%).

To understand more about the phonon mode transport across the interface, we use a Landauer approach to study the diamond-silicon TBC as well. The Landauer approach is a method in frequency space, which facilitates understanding modal phonon transport across the interface compared with NEMD. NEMD simulations include inelastic scatterings naturally from the anharmonic interatomic potentials and could model complicated interface structures, such as an amorphous layer at the interface, while the Landauer approach only considers elastic scatterings and predicts the TBC between bulk materials with perfect interfaces. The two methods provide different insights in the thermal transport across diamond-silicon interfaces so we include both methods here.

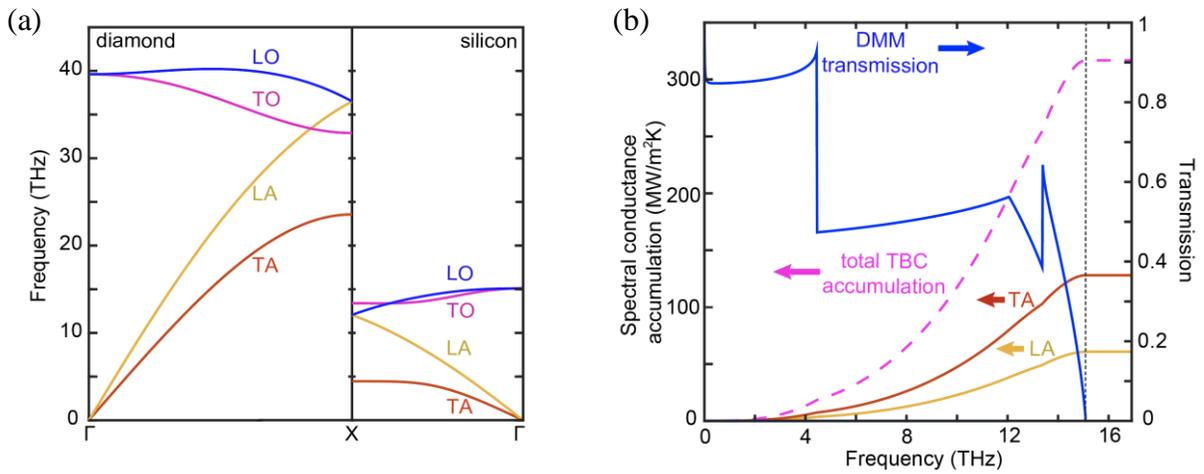

Figure 6. (a) The phonon dispersion relations of silicon and diamond from first-principles calculations. (b) The spectral conductance accumulation and the transmission coefficients from DMM at the interface between diamond and silicon. The left vertical axis is the spectral conductance accumulation while the right vertical axis is the transmission coefficient. The black dotted line is the cutoff frequency of silicon.

The phonon properties of silicon and diamond are calculated from first-principles calculations. The phonon dispersion relation curves, shown in Figure 6(a), are used as inputs to calculate transmission coefficients. Along the Γ–to-X direction in the reciprocal lattice, there are 6 phonon branches: 2 transverse acoustic (TA) branches, 1 longitudinal acoustic (LA) branch, 2 transverse optical (TO) branches, and 1 longitudinal optical (LO) branch. The phonon group velocity (the slope of the dispersion curve) of diamond is much larger than that of silicon, especially for the acoustic branches. The calculated transmission coefficients from DMM are shown in Figure 6(b). In the low frequency range (below 4.5 THz), the number of modes in silicon is much larger than that of diamond. DMM assumes that phonons lose their memory of original directions after reaching the interface. The probability of phonons propagating to the side with larger number of modes is much higher than that to the other side. As a result, the transmission coefficient at low frequency from diamond to silicon is quite high (~0.9). 4.5 THz is the cutoff frequency of the silicon TA branch. Above this frequency, the number of modes on the silicon side decreases sharply so the transmission coefficient drops above this frequency. Here, each turning point in the transmission curve indicates the starting or cutoff frequency of a phonon branch.

The spectral conductance accumulation curve is shown in Figure 6(b). For phonons with frequencies lower than 4 THz, the contribution to TBC is very small because of the small phonon DOS and small phonon energy even though the transmission coefficient is very high. For phonons with higher frequencies, the high spectral contribution to TBC results from the large phonon DOS. The contribution from TA and LA branches to TBC are calculated as well. The contribution from TA branches is twice as that from LA because TA has two branches so the DOS is almost twice as that of LA. The TBC from Landauer is smaller than that from NEMD. We mainly attribute this difference to anharmonic contribution to TBC, which is especially true for diamond-silicon interfaces because the energy diamond phonons could have is much higher than those of silicon phonons. It is possible that multiple silicon phonons scatter at the interface and become one diamond phonon, which contributes to transport energy across the interface (inelastic scatterings).

## 3.2. Enhanced Thermal Conduction in Diamond Membranes

As shown in Figure 2(b), very surprisingly, we find the diamond cross-plane thermal conductivity of the patterned samples grown by graphoepitaxy (Samples A2 and B2) increase by 28% and 10% comparing with that of the flat sample (Sample ref2). To figure out the structure-property relation, we used TEM to study the grain sizes of the diamond layer. Large grains scatter phonons less extensively, leading to a long phonon mean free path and correspondingly high thermal conductivity. In order to measure the grain growth ratio for grains with different orientations, dark field (DF) images were generated over several µm length of the TEM samples. An aperture was used to select the reciprocal lattice points in selected area diffraction patterns that correspond to grains with (111) or (110) planes parallel to the sample surface. The resulting images show the selected grains in bright contrast. As an example, a diamond grain with (110) orientation is shown in the insert of Figure 7(a). The grain width (indicate with red arrows) was measured every 100 nm (as shown with blue arrows) from the depth at which the grain is first observed. We define the "grain growth ratio" as the ratio of the grain size measured at different distances over the grain size measured at 100 nm in order to quantify if grains with certain orientations grow at the expense of others. Figure 7 (a) and (c) show how the grain growth ratios of several diamond crystals with (111) orientation and (110) orientation parallel to the surface change with different distance from the nucleation interface. As depicted in Figure 7, for diamond layers grown on both patterned and flat silicon substrates (Samples A1 and ref1), grains with (111) orientation typically shrink or are blocked by other grains, while grains with (110) orientation tend to expand horizontally while growing. As a result, grains with (110) orientation are longer in the film-thickness (cross-plane) direction than grains with (111) orientation. Similarly, it has been reported that the (110) grain orientation is a preferred grain orientation for CVD diamond growth under certain conditions.[62-64] As discussed above, long grains scatter phonons less extensively in the cross-plane direction, resulting in longer phonon mean free path and correspondingly higher thermal conductivity.

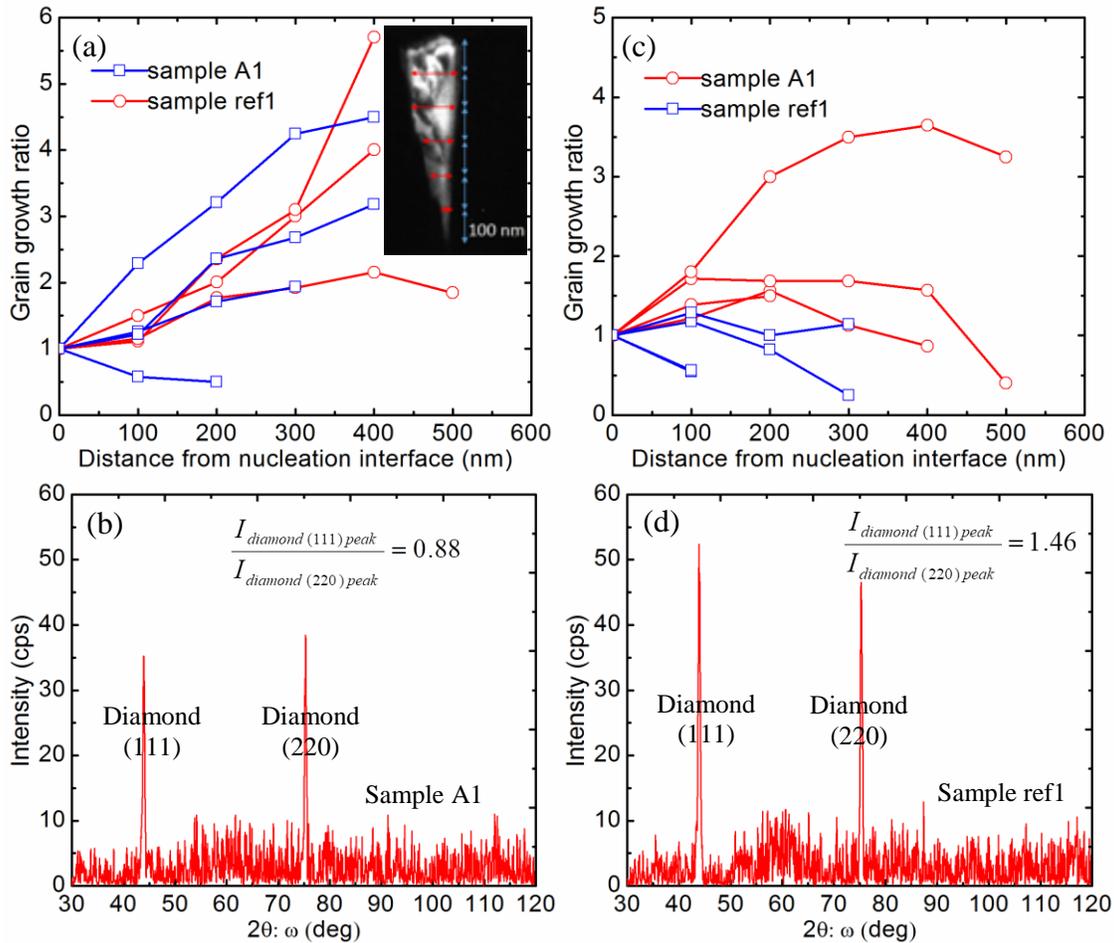

Figure 7. (a) The grain growth ratios of diamond crystals with (110) orientation. The inset (Dark field TEM image to select grains with (110) plane parallel to surface) shows how we measure the grain grown ratio. (b) XRD scan for sample A1. (c) The grain growth ratios of diamond crystals with (111) orientation. (d) XRD scan for sample ref1.

To assess the cross-plane preferred grain orientation, Samples A1, B1, and ref1 were measured using XRD 2θ:ω scans. The XRD peak intensities are from the grains that have that specific plane parallel to the surface and the integrated intensity ratio provides information about texturing. The XRD patterns of Samples A1 and ref1 are shown in Figure 7 (b) and (d) as comparison. The integrated intensity ratio $I_{\text{diamond (111) peak}}/I_{\text{diamond (220) peak}}$ of Samples A1, B1, and ref1 are 0.88, 1.13, and 1.46, respectively. Samples A1 and B1 have smaller integrated intensity ratio than Sample ref1 (all of them are smaller than a ratio of 2.50, assuming no texturing). This feature indicates that all the three samples has (110) texturing while the patterned sample (A1, B1) shows stronger (110) preferred orientation than the flat sample. As discussed above, crystals with (111) orientation typically shrink or are blocked while crystals with (110) orientation are not. When comparing with crystals with (111) orientation, the long crystals with (110) orientation facilitate thermal conduction along the cross-plane direction because of reduced phonon-grain boundary scattering.[65] The higher fraction of grains with (110) orientation in the diamond layer grown by graphoepitaxy indicates higher fraction of long grains, which leads to long phonon mean free path. This result explains the high cross-plane thermal conductivity measured in patterned samples.

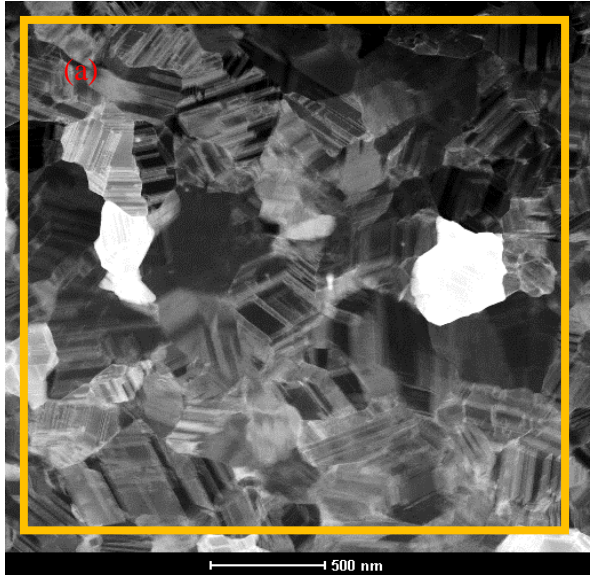

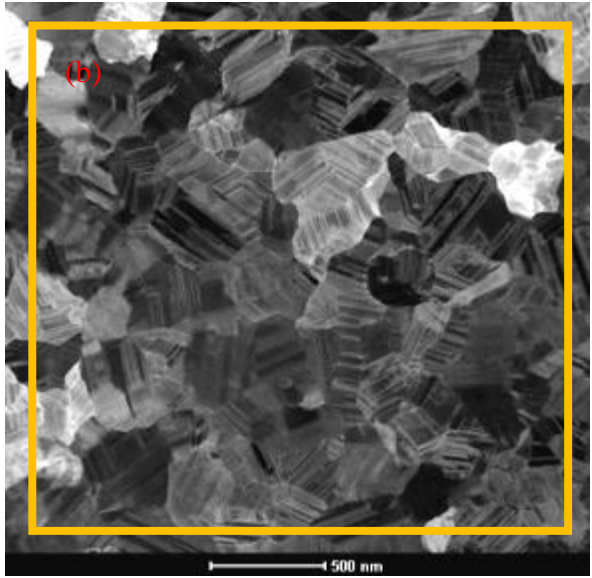

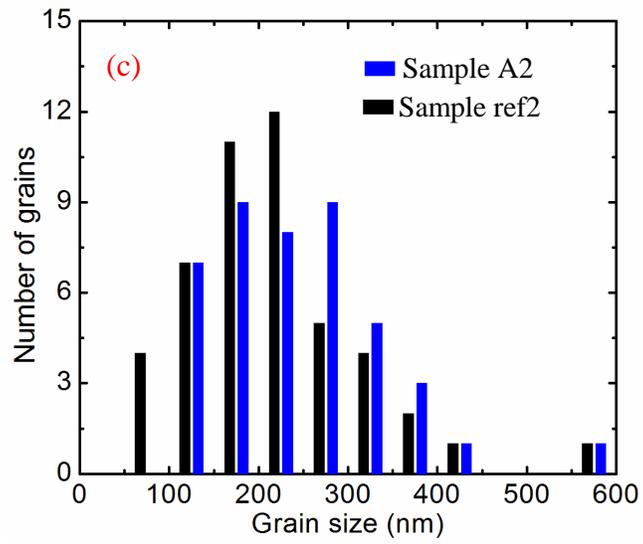

Figure 8. The plan-view STEM image near the diamond film surface for 2 um samples: (a) the patterned sample (Sample A2) and (b) the flat sample (Sample ref2). (c) Grain distribution of Samples A2 and ref2. The average grain size of Sample A2 is 247 nm while that of Sample ref2 is 216 nm.

To further confirm our observations, we also measured the grain distributions of Samples A2 and ref2 with plan-view TEM samples. Figure 8(a-b) show the STEM images of Samples A2 and ref2 near the surfaces of the diamond layers. The grain size is measured within the yellow box and the distribution information is summarized in Figure 8(c). The average grain size of the patterned sample is 247 nm, slightly larger than that of the flat sample (216 nm). Moreover, the patterned sample does not have very small grains (0-100 nm) and has a distribution that is weighted toward larger grain sizes (the patterned sample has 19 grains larger than 250 nm in this area while the flat sample has only 13). Grain boundaries scatter phonons and limit phonon mean free paths, leading to a reduced thermal conductivity.[65-67] The larger average grain size and lower concentration of very small grains scatter phonons less extensively, leading to high thermal conductivity, which supports the observation that the cross-plane thermal conductivity of the diamond grown by graphoepitaxy is higher than that grown on the flat sample.

## 4. CONCLUSIONS

The thermal boundary resistance can be an important factor that limits the heat flow out of high-power-density electronics and microelectronics that require the heterogeneous integration of materials. This is especially true for chemically deposited diamond integrated with other semiconductors due to the large phonon DOS mismatch between diamond and other materials.

However, we show for the first time that it is possible to increase the TBC at semiconductor-dielectric interfaces by graphoepitaxy. By growing diamond on nanopatterned silicon wafers, the present work provides a general strategy to significantly reduce the thermal resistance of both the diamond layer and diamond-substrate interface simultaneously. The diamond-silicon TBC increases by 65% comparing with that of a flat diamond-silicon interface, which is consistent with the contact area enlargement (69%). Our results experimentally confirm the effect of contact area enlargement on TBC predicted by previous theoretical works and achieve the highest diamond-silicon TBC measured to date. The NEMD simulation results show that the amorphous carbon layer at the interface has negligible effect on thermal transport across the interface and the large intrinsic diamond-silicon thermal boundary resistance is the dominant thermal resistance. A Landauer approach is used to calculate diamond-silicon TBC and understand phonon transmissions across the interface. Furthermore, comparing with that of the diamond layer grown on the flat silicon substrate, we observe a 28% increase in thermal conductivity of the diamond layer grown on the patterned substrate which is due to preferred grain orientation (texturing) measured by STEM and XRD. In diamond layers grown on both patterned and flat silicon substrates, grains with (110) orientation typically trend to expand while growing while grains with (111) orientation shrink or are blocked by other grains. XRD results show the diamond layer grown on the patterned substrate has stronger (110) texturing than that on the flat substrate. This finding is confirmed by grain distribution analysis on diamond grain sizes near the grown side for Samples A2 and ref2. The average grain size of the patterned sample A2 (247 nm) is slightly larger than that of the flat sample ref2 (216 nm). Moreover, the patterned sample does not have very small grains (0-100 nm) and has a distribution that is weighted toward larger grain sizes. Graphoepitaxy provides a general solution to significantly enhance thermal transport across diamond layers and diamond-substrate interfaces when integrating diamond to substrates for applications of electronics cooling.

## SUPPORTING INFORMATION

SEM pictures of nanoscale patterns (Figure S1), TDTR sensitivity (Figure S2) analysis and data fitting (Figure S3), TEM sample preparation instruction with FIB (Figure S4).

## ACKNOWLEDGEMENTS

The authors would like to acknowledge the funding support from U.S. Defense Advanced Research Projects Agency (DARPA) Diamond Round Robin Program "Thermal Transport in Diamond Thin Films for Electronic Thermal Management" under contract no. FA8650-15C. The NEMD simulations by T.L.F. and S.T.P. are supported in part by Department of Energy grant DE-FG0209ER46554 and by the McMinn Endowment. Computations at Vanderbilt University and ORNL were performed at the National Energy Research Scientific Computing Center (NERSC), a Department of Energy, Office of Science, User Facility funded through Contract No. DE-AC02-05CH11231. Computations also used the Extreme Science and Engineering Discovery Environment (XSEDE).

## COMPETING FINANCIAL INTEREST

A patent related to this work has been applied with application number 62/676,294 filed by May 25, 2018.

## ORCID

Samuel Graham: 0000-0002-1299-1636

## Graphical Table of Contents

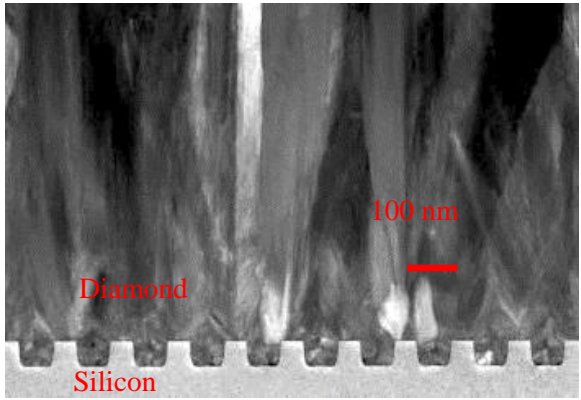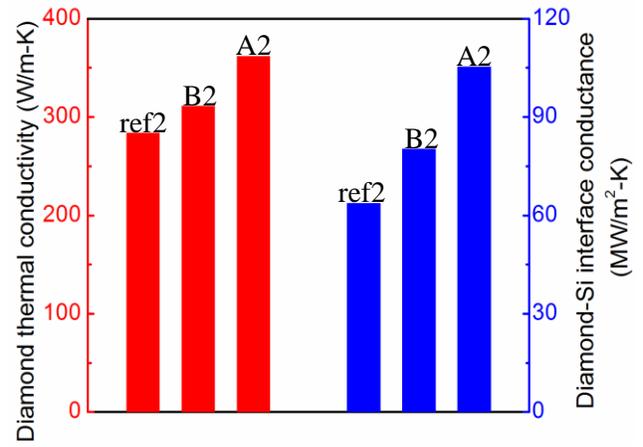